\documentstyle[11pt,newpasp,twoside,epsf]{article}
\def\edcomment#1{\iffalse\marginpar{\raggedright\sl#1\/}\else\relax\fi}
\reversemarginpar

\begin{document}
\title{The Great Austral Nearby Young Association \altaffilmark {1} }
\altaffiltext {1}{Based on observations made under the Observat\'{o}rio Nacional-ESO agreement for the joint operation of the 1.52\,m ESO telescope and at the  Observat\'{o}rio do Pico dos Dias, operated by MCT/Laborat\'{o}rio Nacional de Astrof\'{\i}sica,  Brazil} 
\author{ C. A. O. Torres, G.R. Quast} 
\affil{Laborat\'{o}rio Nacional de Astrof\'{\i}sica/MCT, 37504-364 
Itajub\'{a}, Brazil}
\author{R. de la Reza,  L. da Silva}
\affil{Observat\'{o}rio Nacional/MCT, 20921-030 Rio de Janeiro, Brazil}
\author{C. H. F. Melo}
\affil{Geneva Observatory, CH-1290 Sauverny, Switzerland}
%\email{germano@lna.br, beto@lna.br, ldasilva@eso.org, delareza@on.br, Claudio.Melo@obs.unige.ch}

\begin{abstract}
Observing ROSAT sources in an area covering $\sim$30\% of the Southern Hemisphere, we found evidences for a great nearby association (GAYA), comprising the proposed associations of  Horologium (HorA) and the eastern part of Tucana
(TucA), formed by at least 44 Post-T\,Tauri stars.
The stars of the GAYA have similar space 
velocity components relative to the Sun 
\mbox{(U, V, W)\,=\,($-9.8 \pm 1.2, -21.7 \pm 1.1, -2.0 \pm 2.2$)\,km/s}
and their Li line intensities are between those of the classical T\,Tauri stars and the ones of the Local Association stars. 
The distances of the members of the GAYA  cover an interval of $\sim$70\,pc,
compatible with the angular size of $\sim60\deg$  and in agreement with 
an initial velocity dispersion of $\sim$1.5\,km/s and its evolutive age.  
We found  many other  young stars, not members of  the GAYA.
We also observed a control region near the equator, covering 700 square degrees,  where we found only four young stars.  
The overabundance of young stars  near the South Pole (by a factor of five) 
seems to show that there may be other young associations not yet characterized. 
\end{abstract}

\section{Introduction}
\begin{figure}
\plotone{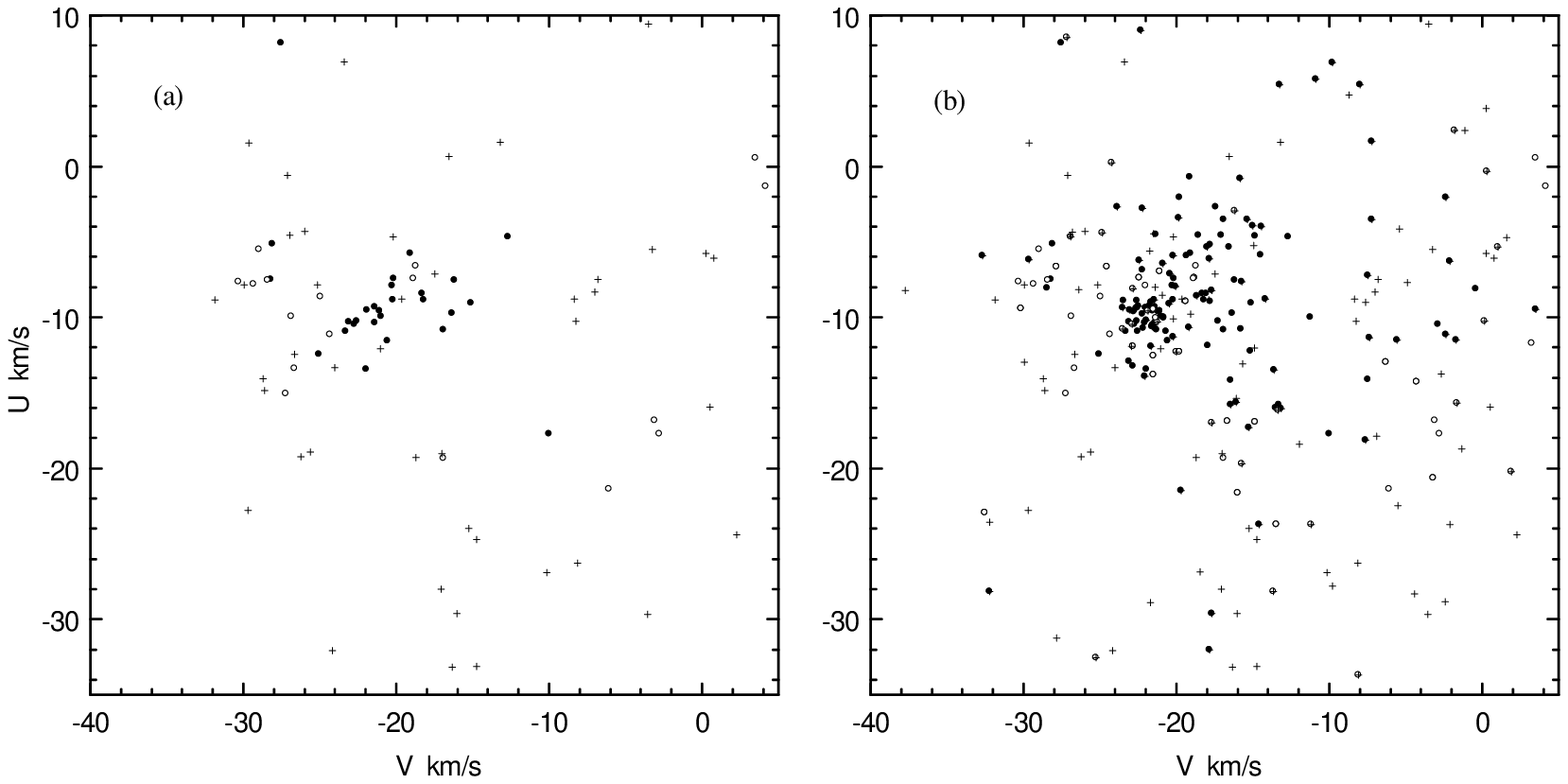}
\caption{UV space velocities for the stars observed in the SACY.\\ 
(a) Hipparcos stars.\\ 
(b) All stars, including those with kinematical parallaxes.\\ 
Filled circles, open circles and plus  represent the stars with 
Li\,I lines stronger, similar or weaker (or absent) 
than the Pleiades, respectively.  
The concentration near (-10, -20), in both figures, is the GAYA. 
There are stars out of the figure boundaries.}
\end{figure}

\begin{figure}
\plotone{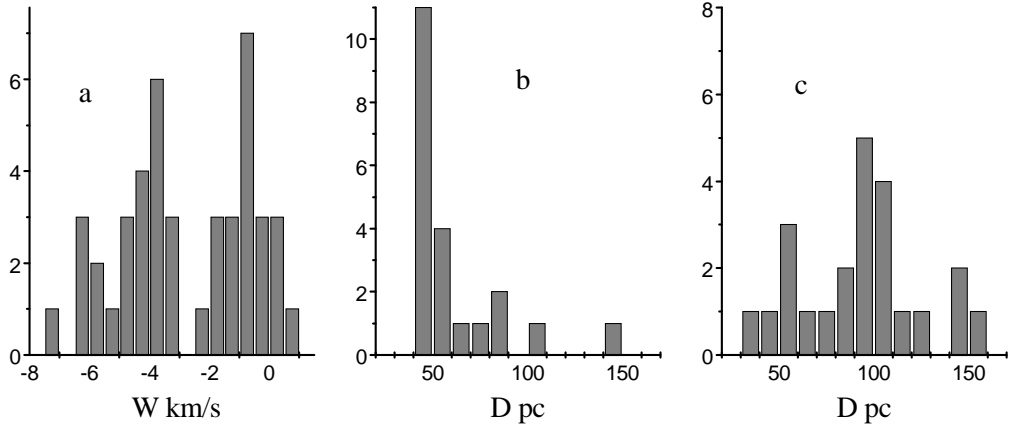}
\caption{(a) is the histogram of the W velocites for the proposed members
of the GAYA. 
There seems to be a bimodal distribution that reflects in the
distribution of the distances: 
(b) for the right peak and (c) for the left one. 
It is not yet clear if these distributions represent a real 
split in the GAYA or some bias in the sample or introduced in some way 
by our statistical approach.}
\end{figure}

Surveys of young pre-main-sequence (PMS) stars based on IRAS colors 
detect mainly the classical T\,Tauri stars (CTT), 
due to their important dusty accretion disks. 
One of the most comprehensive surveys was the ``Pico dos Dias Survey'' (PDS) (Gregorio-Hetem et al. 1992; Torres et al. 1995; Torres 1998).  
As the disks around the Post-T\,Tauri stars (PTT) should have been 
dissipated or agglutinated  into planetesimals, 
they hardly  would be detected in  this way. 
They could belong to physically dispersed groups, with ages larger than 
the mean lifetime of the  original clouds, and, in such case, 
they may be located far from any cloud. 
We discovered the first of this kind of 
association when we were searching for new T\,Tauri stars (TTS) around 
TW\,Hya (de la Reza et al. 1989; Gregorio-Hetem et al. 1992).  
The detection of X-ray sources  by the ROSAT All-Sky Survey (RASS)
associated with TTS outside star formation 
regions (Neuh\"{a}user 1997) gave a new tool to find 
other associations of this kind. 
In fact, Torres et al. (2000), and Zuckerman \& Webb (2000), 
using these sources, found evidences for two young associations 
near the South Pole, the Horologium (HorA)  and the Tucana (TucA) Associations. 
To examine the possibility of these associations being the same and
to search for other ones we undertook a Survey for Associations Containing
Young-stars (SACY).

\section{Results}

We observed more than 400 stars in the surveyed area defined in de la Reza 
et al. in these proceedings. 
We obtained spectral classifications, radial velocities and equivalent width of Li\,I lines.
The Li\,I line in late type stars can provide a first estimate for the age (Jeffries, 1995), selecting possible PTT. 
In SACY we consider that a star is probably younger than the 
Local Association if  it is located near or above the 
limit proposed by Neuh\"{a}user (1997).

In Figure\,1a we plot the (U, V) space motions of the stars with 
Hipparcos parallaxes. There is a concentration of Li-rich stars near the
position (-10, -20). Choosing the 10 stars near this position (visual binaries
taken as mean values) we obtain as the mean space velocity components:\\
\centerline{(U, V, W)\,=\,($-9.8 \pm 1.2, -21.7 \pm 1.1, -2.0 \pm 2.2$)\,km/s}
The mean parallax of the stars is $20.2\,\pm\,2.2$\,mas. 
We used the kinematical method, discribed in Torres et al. (2000), to estimate the  distances and space velocities for the stars not measured by Hipparcos.

In Figure\,1b we plot the space motions, deduced in this way, 
of the observed stars. 
There is a compact core of 44 Li strong stars, which are the 
``probable members'' of a Great Austral Young Association (GAYA). 
In Figure\,2a we show the histogram of the W velocities. 
It shows a double peak as if there are two similar associations.
In Figure\,2b and 2c we show the historgrams of the distances of both
groups. The group with greater W seems closer than the other.
Is the GAYA actually split in two associations?

We made a global analysis with all stars having Li line 
similar to or stronger than the Pleiades, 
computing a grid of convergence points in velocity space. 
The above two groups stand out as the strongest concentrations. 
We tested also the convergence for the space velocities of the Chamaeleon  
complex and obtained only a diffuse and little convincing concentration
(as the possible stars are spread in all the observed area). 
Only four stars of  the GAYA could also belong to the Chamaleon, 
but none is in its neighborhood.

In Figure\,3 we show the celestial map of the observed stars. 
Those not in the concentration in the UV plane are 
evenly distributed, even the possible young ones. 
The proposed members of the GAYA are concentrated ``near'' 
the South Pole, between RA\,=\,21H and 9H. 
The GAYA includes the HorA and the eastern part of TucA. 
It is so large that it is meaningless to call it for any constellation.
The GAYA seems also to split in the sky, one of the groups being more concentrated 
to the east.
But as its boudaries are not yet clearly established, 
and as in all diagrams and figures there is some superposition, 
we must wait for more observations to confirm this split. 
Anyway, there is no clear evolutionary distinction between the groups,
both being very similar, having $\sim$30\,Myr.
\begin{figure}

\plotfiddle{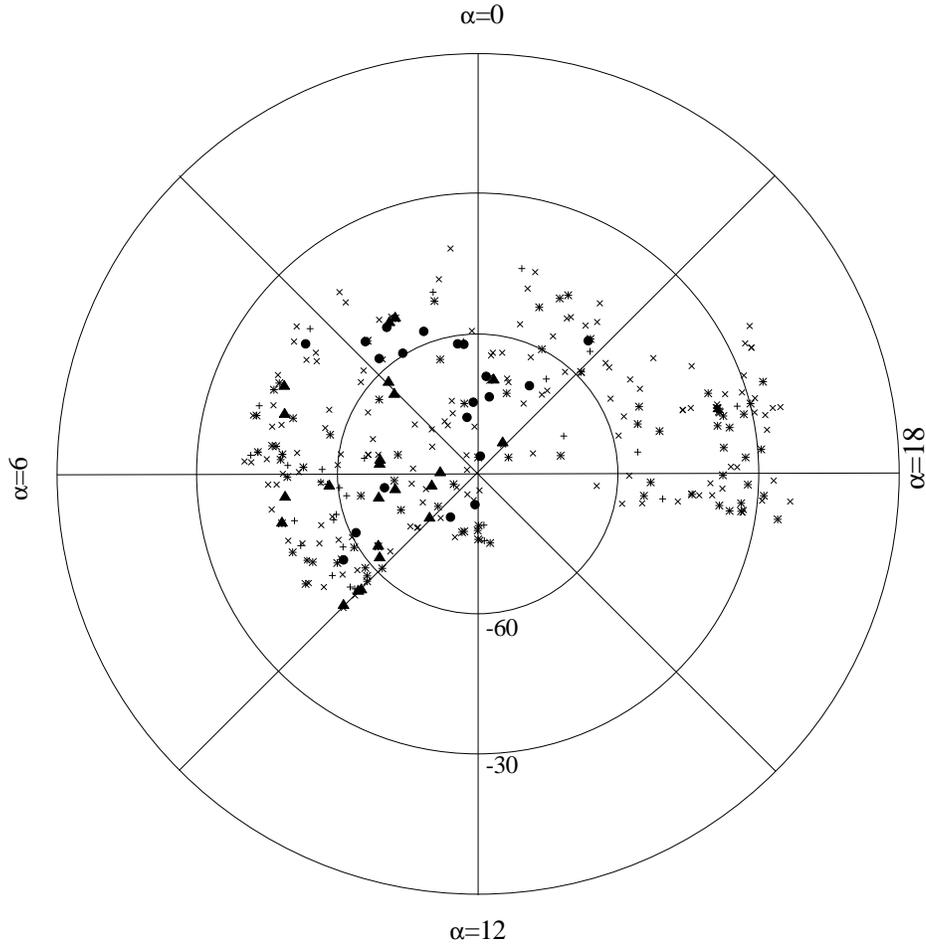}{11cm}{0}{70}{70}{-215}{-210}
%\plotone{torres_c_fig3.eps}
\vspace{20pt}
\caption{Celestial polar projection of the stars observed in SACY (giants excluded). 
Circles are the GAYA members with higher W (Fig\,2b) 
and triangles with lower W (Fig\,2c). 
Other symbols are for non-members: 
asterisks, plus and crosses for stars with Li lines stronger, similar or weaker (or absent) than the Pleiades, respectively.}
\end{figure}

The  kinematical parallaxes give larger distances ($\sim$150\,pc) 
for some stars but they are unreliable.
Thus, the depth of the GAYA, from Figure\,2, is $\sim$70\,pc,  
implying a projected  angular diameter of $\sim60\deg$, 
similar to the observed size. 
If we suppose that the original velocity dispersion  during star formation 
is  equal to the average modulus of the velocity vectors  ($\sim$1.5\,km/s),  
the age of the GAYA would be $\sim$20\,Myr,
similar to the evolutive age.

For most of the  probable  GAYA members we obtained at least three
radial velocities  with no indication of 
spectroscopic binaries. 
This possible low binary frequency is intriguing
and in contrast with  the TWA case.

The surveyed area has many other young stars and 
they should be investigated for other possible associations,
even if the global analysis does not show any other proeminent concentration.
In fact, the high austral density of young stars seems to be real, 
being five times higher than the observed one in the region around BP\,Psc, 
where in 700\,square degrees we found only four young stars.

\section {Conclusions}

Exploring a region covering  about 30\% of the Southern Hemisphere 
we  found  near the South Pole an association of 
at least 44 young stars (GAYA), 
younger (age $\sim$30\,Myr) than most groups of the Local Association 
and older than the TWA,  ingulfing the HorA and part of the TucA.
We did not yet detect spectroscopic binaries in the GAYA. 
The distances of the members of the GAYA  cover an interval of $\sim$70\,pc,
compatible with the size produced after 20\,Myr  by  an initial velocity dispersion of $\sim$1.5\,km/s, in agreement with the evolutive age.

The GAYA is formed only by  PTT, where none is an IRAS source, 
indicating the absence of dusty accretion disks.
This suggests that, at their advanced evolutionary stage, 
the circumstellar disks have been dissipated 
or agglomerated into planetesimals.    
Thus, the members of the GAYA are privileged targets for studying 
the formation and early evolution of planets.

\acknowledgments

C. A. Torres thanks FAPEMIG, G. R. Quast CNPq and R. de la Reza CAPES  for providing financial support.
This work was partially supported by a CNPq grant to L. da Silva 
(pr. 200580/97) and to C. Melo (proc. 200614/96-7).
We thank M. Sterzik and B. Reipurth for instructive discussions 
and M. Mayor for the use of the Swiss telescope.

\end{document}